# Long-term stability of phase-separated Half-Heusler compounds

Julia Krez, Benjamin Balke[*], Claudia Felser, Wilfried Hermes and Markus Schwind

*Dr. J. Krez, Dr. B. Balke, Institute for Inorganic and Analytical Chemistry, Johannes Gutenberg University Mainz, Staudingerweg 9, 55128 Mainz, Germany;*

[*]*Dr. B. Balke, Email: balke@uni-mainz.de*

*Dr. J. Krez, Graduate School for Excellence Materials Science in Mainz, Johannes Gutenberg University Mainz, Staudingerweg 9, 55128 Mainz, Germany*

*Prof. Dr. C. Felser, Max Planck Institute for Chemical Physics of Solids, Nöthnitzer Str. 40, 01187 Dresden, Germany*

*Dr. Wilfried Hermes, Dr. Markus Schwind, BASF SE, 67056 Ludwigshafen, Germany*

Half-Heusler (HH) compounds have shown high figure of merits up to 1.5[1,2,3]. The key to these high thermoelectric efficiencies is an intrinsic phase separation[4], which occurs in multicomponent Half-heusler compounds and leads to an significantly reduction of the thermal conductivity. For commercial applications, compatible *n*- and *p*-type materials are essential and their thermal stability under operating conditions, e.g. for an automotive up to 873 K[5], needs to be guaranteed. For the first time, the long-term stability of *n*- and *p*-type HH materials is proved. We investigated HH materials based on the $Ti_{0.3}Zr_{0.35}Hf_{0.35}NiSn$-system after 500 cycles (1700 h) from 373 to 873 K. Both compounds exhibit a maximum Seebeck coefficient of $|\alpha| \approx 210$ µV/K and an intrinsic phase separation into two HH phases. The dendritic microstructure is temperature resistant and maintained the low thermal conductivity values ($\kappa < 4$ W/Km). Our results emphasize that phase-separated HH compounds are suitable low cost materials and can lead to enhanced thermoelectric efficiencies beyond the set benchmark for industrial applications.

## Introduction

The search for alternative energy technologies has intensified in recent years as climate change has become more noticeable and the usage of nuclear energy introduces political controversy for many countries. The challenge here lies in developing sustainable energy conversion techniques that can compete with the efficiency of fossil fuels. In terms of waste heat recovery, thermoelectric generators (TEG) provide an excellent method for generating electricity from dissipated heat[6,7]. Especially in the automobile industry, TEGs play a crucial role in decreasing greenhouse gas emissions and fuel consumption[8]. For commercial and versatile application of TEG in power generation, thermoelectric (TE) materials need to meet several criteria besides high efficiency, such as environmentally friendliness, low cost production and long-term thermal stability (see Figure 1).

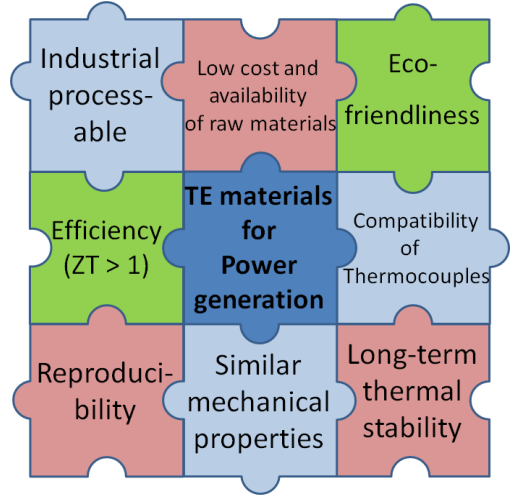

**Figure 1.** Illustration of the criteria for TE materials required for large-scale power generation.

The TE efficiency of a given material is determined by the dimensionless figure of merit $zT$ as given in Equation (1):

$$zT = \frac{\alpha^2 \sigma}{\kappa} T, \quad (1)$$

where $\alpha$ is the Seebeck coefficient, $\sigma$ the electrical conductivity, and $\kappa$ the total thermal conductivity[6]. *Slack* introduced the "phonon-glass electron crystal" concept and showed that there is a limitation of $zT$, since the required materials properties, such as high electrical conductivity and low thermal conductivity, are strongly related to each other[9]. The general strategy towards high TE efficiency can be attained by enhancing the electronic properties through optimization of the carrier concentration and by reducing the thermal conductivity through effective phonon scattering[6,7].

Most current research on high-$zT$ TE materials is focused on introducing nanostructures or nanoinclusions[10,11]; however, it is doubtful whether the criteria of reproducibility and thermal stability are met by nanostructured materials especially under operating at high temperatures[12]. Intermetallic HH compounds with a general formula *XYZ* (*X, Y* = transition metals; *Z* = main group element like Sn or Sb) have attracted considerable attention as suitable TE materials because of their very flexible electronic structure (band gap tunability from zero to 4 eV)[13] and high Seebeck coefficients ($|\alpha| > 200$ µV/K)[11] (see Supplementary Information Figure S1 and S2). Thereby, the transport properties can easily be changed from *n*-type to *p*-type behavior by electronic doping with non-isoelectronic elements[11,14]. Thus, donor impurities (elements with higher valence electron count) lead to a shift of the Fermi energy towards the conduction band (higher energies), while acceptor impurities (elements with higher valence electron count) will shift the Fermi energy into the valence band (lower energies). Furthermore, multi-component HH compounds exhibit an intrinsic phase separation during their solidification process[15]. This eutectic microstructure is extraordinarily temperature stable, owing to the high melting points (> 1400 K)[16] of HH materials. Temperature stable eutectic microstructures can establish an entirely new area of research, where the phase-separation microstructure can be independently engineered to fulfill desired TE properties[4], leading to further improvements in TE efficiencies beyond the current state-of-the-art materials.

Numerous studies investigated the TE properties of multi-component *M*NiSn- and *M*CoSb-based HH materials[3,11,17]. Thereby, *zT* values up to 1.5 for *n*-type materials[1] and 1.0 for *p*-type materials[3] could be attained. However, to the best of our knowledge, there are no reports on the importance of the long-term stability of HH compounds or the preservation of their TE properties through repeated heating and cooling. Very recently, *Bartholomé et al*[18] investigated the reproducible up-scaling process of *n*-type $Zr_{0.4}Hf_{0.6}NiSn_{0.98}Sb_{0.02}$ and *p*-type $Zr_{0.5}Hf_{0.5}CoSb_{0.8}Sn_{0.2}$ into kilogram batches. The TE module comprising these HH compounds contained 7 *n*- and 7 *p*-type legs and exhibited a maximum power output of 2.8 W. This work presents the first study on the long-term stability of phase-separated (Ti,Zr,Hf)NiSn HH compounds. The exemplary *n*-type $Ti_{0.3}Zr_{0.35}Hf_{0.35}NiSn$ and *p*-type $Ti_{0.26}Sc_{0.04}Zr_{0.35}Hf_{0.35}NiSn$ HH compounds were subjected to 500 heating-cooling cycles from 373 to 873 K, lasting about 1700 h. The TE properties were measured after 50, 100, and 500 cycles up to 900 K.

**Results**

The crystalline structure of the compounds was determined by powder *X*-ray diffraction (PXRD) measurements. The obtained *X*-ray patterns reveal a cubic C1$_b$ structure (F$\bar{4}$3m), which is in agreement with results found in the literature[19]. Splitting of the Bragg reflection peaks is caused by the coexistence of two

HH phases. During the solidification process, the multi-component HH system dissipates into a Ti-poor (HH 1) and a Ti-rich (HH 2) phases[15]. The lattice parameters of both *n*-type $Ti_{0.3}Zr_{0.35}Hf_{0.35}NiSn$ and *p*-type $Ti_{0.26}Sc_{0.04}Zr_{0.35}Hf_{0.35}NiSn$ HH compounds are listed in Table 1. According to the 4% Sc substitution and larger radius of Sc (Ti 140 pm, Sc 160 pm), the lattice constant is higher for the *p*-type material, indicating that Sc substituted into the Ti lattice.

**Table 1.** Lattice parameters and chemical compositions determined using energy dispersive *X*-ray analysis of phase-separated *n*-type $Ti_{0.3}Zr_{0.35}Hf_{0.35}NiSn$ and *p*-type $Ti_{0.26}Sc_{0.04}Zr_{0.35}Hf_{0.35}NiSn$ HH compounds.

|  | Lattice parameter *a* [Å] | Composition |
|---|---|---|
| $Ti_{0.3}Zr_{0.35}Hf_{0.35}NiSn$ (*n*-type) | | |
| Ti-poor phase (HH 1) | 6.068(2) | $Ti_{0.18}Zr_{0.40}Hf_{0.41}NiSn$ |
| Ti-rich phase (HH 2) | 6.012(3) | $Ti_{0.65}Zr_{0.2}Hf_{0.17}NiSn$ |
| $Ti_{0.26}Sc_{0.04}Zr_{0.35}Hf_{0.35}NiSn$ (*p*-type) | | |
| Ti-poor phase (HH 1) | 6.071(7) | $Ti_{0.14}Sc_{0.04}Zr_{0.4}Hf_{0.41}NiSn$ |
| Ti-rich phase (HH 2) | 6.045(6) | $Ti_{0.62}Sc_{0.04}Zr_{0.20}Hf_{0.17}NiSn$ |

The backscattering electron mode (BSE) SEM images of the *n*- and *p*-type samples before and after 500 cycles, shown in Figure 2, revealed the intrinsic phase separation into the main Ti-poor HH 1 phase and the Ti-rich HH 2 phases, which is dendritically interlaced through the microstructure. The compositions of the phases for the *n*- and *p*-type HH compounds are listed in Table 1. Since this (Ti,Zr,Hf)NiSn HH system exhibits a melting point of ≈1720 K (see Supplementary Information Figure S3), the resulting phase separation is stable under the cycling conditions, which is of utmost importance for maintaining low thermal conductivities in these HH materials.

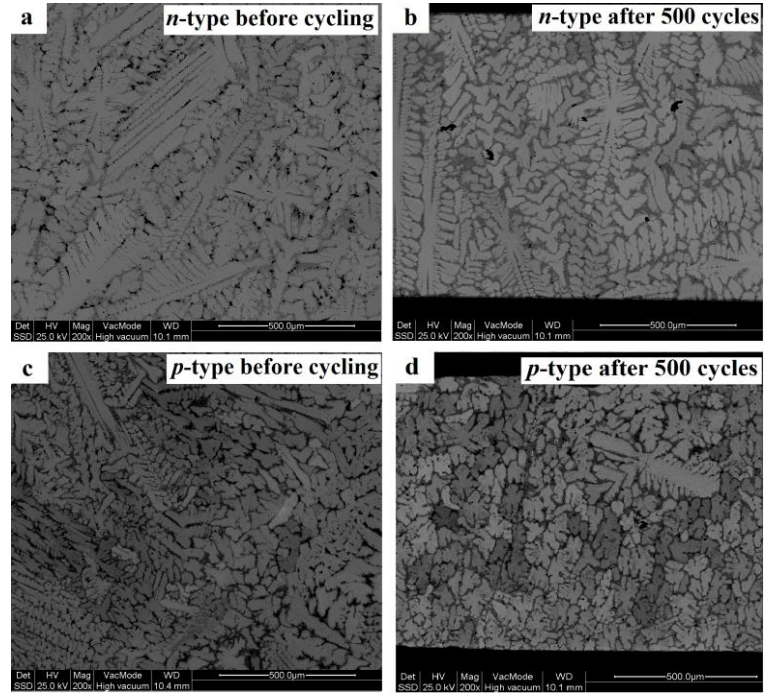

**Figure 2.** SEM images of the *n*-type $Ti_{0.3}Zr_{0.35}Hf_{0.35}NiSn$ and *p*-type $Ti_{0.26}Sc_{0.04}Zr_{0.35}Hf_{0.35}NiSn$ HH compound. a) *n*-type compound before cycling. b) *n*-type compound after 500 cycles. c) *p*-type compound before cycling. d) *p*-type compound after 500 cycles. All compounds reveal an intrinsic phase separation into a Ti-poor HH 1 phase (grey) and Ti-rich HH 2 (dark) phase.

The transport properties of the semiconducting *n*-type $Ti_{0.3}Zr_{0.35}Hf_{0.35}NiSn$ HH system were measured up to 873 K after 500 cycles (1700 h) (Figure 3). The electrical conductivity $\sigma$ improved after 50 cycles due to an enhancement of the structural order (Figure 3a)[11]. The Seebeck coefficient $\alpha$ is negative at all temperatures, indicating electrons as the majority charge carriers, with a peak value of $\alpha \approx$ -210 µV/K at 600 K (Figure 3b). The large Seebeck coefficient emerges from the high density of states, which is caused by the *d*-states of the transition metals near the Fermi level[19]. The decrease in $\alpha$ above 600 K is caused by the thermal excitation of intrinsic carries. The power factor *PF* shows a peak value of $2.5*10^{-3}$ W/Km at 773 K after 50 cycles (see Supplementary Information Figure S4), which is comparable to values of *n*-type $Bi_2Te_3$, i.e., $PF \sim 2.6*10^{-3}$ W/Km at 423 K[20]. A recent investigation of the long-term efficiency of a commercial available bulk-$Bi_2Te_3$ TEG with 31 thermocouples showed a reduction in $\alpha$ and $\sigma$ caused by material deterioration[21]. The thermal conductivity $\kappa$ of (Ti,Zr,Hf)NiSn compounds is much lower than of ternary HH alloys, where $\kappa > $ 7–17 W/Km[22]. The low $\kappa$ values of the phase-separated HH system occur due to different phonon scattering agents in the structure. The impedances of high-frequency phonons are induced by alloy scattering, while low and mid-frequency phonons are effectively scattered at the dendritic interfaces[23]. Owing to the improved structural order of the *n*-type compound, $\kappa$ increases slightly after 500 cycles (Figure 3c). The increase in $\kappa$ above 600 K is caused by the excitation of intrinsic carriers. The figure of merit, *zT* shows no drastic change after 500 cycles since the values lie within the acceptable error range (Figure 3d).

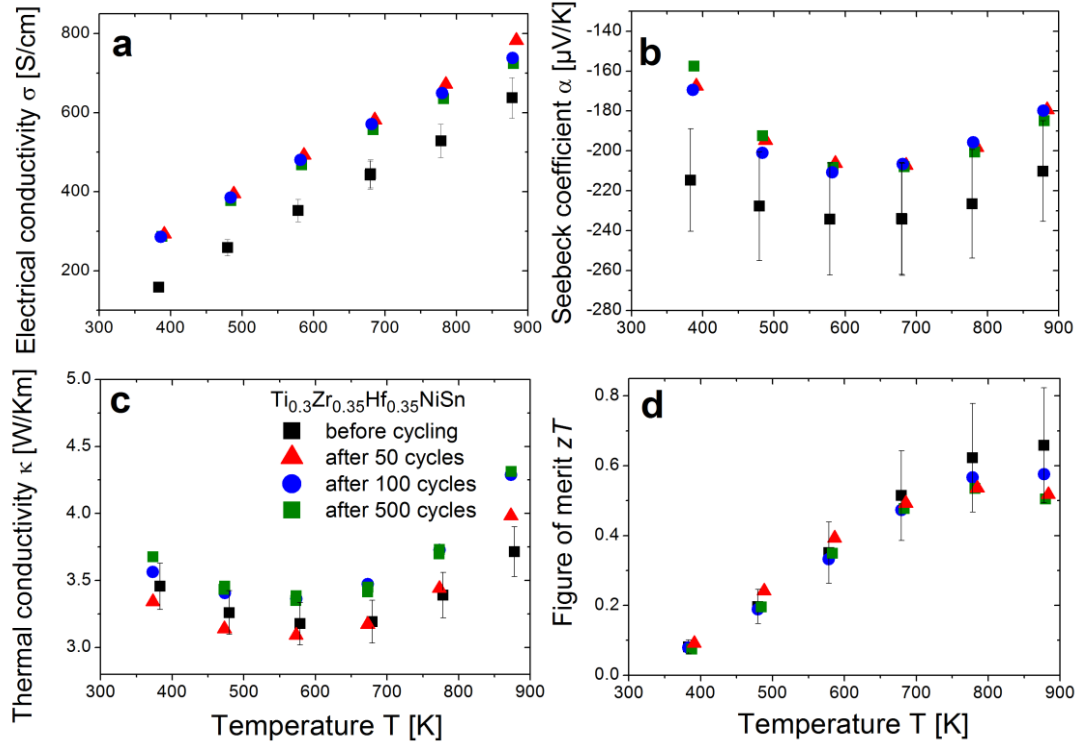

**Figure 3.** Thermoelectric properties as a function of temperature of the *n*-type $Ti_{0.3}Zr_{0.35}Hf_{0.35}NiSn$ compound under the long-term treatment for 500 cycles. a) Electrical conductivity $\boldsymbol{\sigma(T)}$ with an error of 8%. b) Seebeck coefficient $\boldsymbol{\alpha(T)}$ with an error of 11%. c) Thermal conductivity $\boldsymbol{\kappa(T)}$ with an error of 5%. d) Figure of merit $\boldsymbol{zT(T)}$ with an error of 25% under consideration of the measurement uncertainty and error propagation.

The substitution of 4% Sc on the (Ti,Zr,Hf)-site in the *n*-type $Ti_{0.3}Zr_{0.35}Hf_{0.35}NiSn$ parent compound leads acceptor impurities, shifting the Fermi level into the valence band, resulting in *p*-type conductivity for the $Ti_{0.26}Sc_{0.04}Zr_{0.35}Hf_{0.35}NiSn$ compound. The temperature dependence of the electrical conductivity $\sigma$ shows semiconducting behavior (Figure 4a). The $\sigma$ values for the *p*-type are lower than those for the *n*-type, caused by the diminished mobility of carriers due to the shift of the Fermi level into the heavier valence band[6]. The Seebeck coefficient $\alpha$ is positive over the entire temperature range, reaching a maximum of $\alpha \sim 210$ µV/K at 600 K (Figure 4b), which is the highest reported value for a *p*-type material based on a (Ti,Zr,Hf)NiSn system[19]. Both *n*- and *p*-type HH compounds exhibit equally high Seebeck coefficients within the same temperature range. The decrease in the absolute Seebeck coefficient $\alpha$ for both systems above 600 K arises from the onset of minority carrier activation. The thermal conductivity $\kappa$ is further reduced for the *p*-type compound ($\kappa \sim$ 2.5 mW/K) below 600 K (Figure 4c), owing to alloy scattering resulting from the 4% Sc substitution[24]. It can be seen in Figure 4d, that the TE properties exhibit only a weak dependence on cycling conditions (1700h).

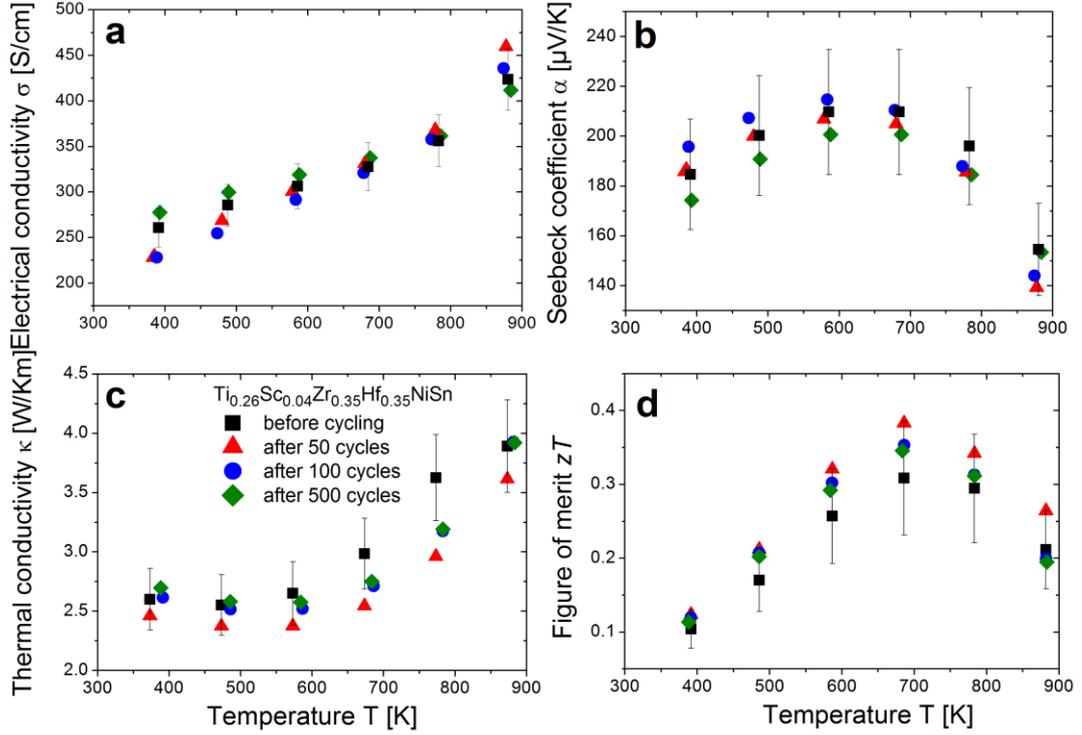

**Figure 4.** Thermoelectric properties as a function of temperature of the *p*-type $Ti_{0.26}Sc_{0.04}Zr_{0.35}Hf_{0.35}NiSn$ compound under the long-term treatment for 500 cycles. a) Electrical conductivity $\sigma(T)$ with an error of 8%. b) Seebeck coefficient $\alpha(T)$ with an error of 11%. c) Thermal conductivity $\kappa(T)$ with an error of 5%. d) Figure of merit $zT(T)$ with an error of 25% under consideration of the measurement uncertainty and error propagation.

We were able to prove the long-term stability for two compatible HH materials, the *n*-type $Ti_{0.3}Zr_{0.35}Hf_{0.35}NiSn$ and *p*-type $Ti_{0.26}Sc_{0.04}Zr_{0.35}Hf_{0.35}NiSn$. Their properties are stable even after 500 cycles (1700 h) in a temperature range from 373 to 873 K. The figure of merit $zT$ for both *n*- and *p*-type compounds does not change significantly under the long-term treatment. The dendritic microstructure consisting of a Ti-poor HH 1 and a Ti-rich HH 2 phase in both *n*- and *p*-type HHs is stable under the long-term cycling, which is crucial to maintain the low thermal conductivities in the HH materials. This approach involving an induced thermally stable eutectic microstructure could be the key for improving the thermoelectric performances of HH compounds above known limits. Our results strongly demonstrate the suitability of phase-separated HH materials, which also comply with requirements such as reproducibility and environmental friendliness via mechanical and thermal stability, for a commercial TE application at moderate temperature.

Experimental Section

*Materials Synthesis*: The samples $Ti_{0.3-x}Sc_xZr_{0.35}Hf_{0.35}NiSn$ (x = 0.04) were prepared by arc melting of Ti, Sc, Zr, Hf, Ni, Sn (purity of all elements was 99,9%) in a purified argon atmosphere using a mixture of stoichiometric amounts of these elements. To ensure homogeneity, the samples were re-melted several times. The as-cast samples were annealed at 1223 K for 7 days in order to improve the crystalline order, and this was followed by quenching in ice water. The samples were cut into rectangular bars and subjected to a temperature profile from room temperature up to 873 K. Fifty cycles of temperature increase/decrease lasted approximately 170 h.

*Materials characterization*: XRD measurements were performed on a Bruker D500 diffractometer with Cu Kα radiation. The lattice parameter was determined using $LaB_6$ as internal standard. The chemical compositions of the samples with ∼ 0.5–0.3 at% were examined by scanning electron microscopy (SEM) using a Philips X scanning electron microscope. Quantitative electron probe microanalysis (EPMA) of the phases was carried out by using a wavelength dispersive *X*-ray analyzer (Phoenix V 5.29, EDAX) and a wavelength-dispersive spectrometer (Cameca SX 100) with the pure elements as standards (the acceleration voltage was 25 kV, using the *K*- and *L*-lines). The electrical conductivity σ and Seebeck coefficient $\alpha$ were measured simultaneously under a helium gas atmosphere from room temperature up to about 900 K using a Linseis LSR-3 system. Heating and cooling cycles yielded repeatable electrical properties for a given sample. The thermal conductivity κ was evaluated as a product of the thermal diffusivity $D$, heat capacity $c_p$, and measured density $\rho$. The thermal diffusivity $D$ was measured by a laser flash method using the Netzsch LFA 457. All samples were coated with a thin layer of graphite to minimize emissivity errors. The heat capacity $c_p$ was measured by differential scanning calorimetry using the Netzsch STA 449. The density was determined via a geometrical method by measuring the weight and dimensions of the rectangular bars before and after the thermal treatment.

Acknowledgements:

We gratefully acknowledge the German BMBF joint project TEG 2020 for their funding. And we would like to thank J. Schmitt and U. Burkhardt for many helpful discussions, R. Stinshoff for the electron microscope measurements.

[1] S. Sakurada, N. Shutoh, *Appl. Phys. Lett.* **2005**, 86, 082105.

[2] M. Schwall, B. Balke, *Phys. Chem. Chem. Phys.* **2013**, 15, 1868.

[3] X. Yan, W. Liu, S. Chen, *Adv. Energy. Mater*. **2013**, 3, 1195.

[4] M. Koehne, T. Graf, H.-J. Elmers, C. Felser *US patent 0156636,* **2013,** Jun. 20.

[5] K. Schierle-Arndt and W. Hermes, *Chem. Unserer Zeit,* 2013, **47**, 92-101.

[6] G.J Synder, E.S. Toberer, *Nat Mater.* **2008**, **7**, 105.

[7] T.M Tritt, *Annu. Rev. Mater. Res*. **2011**, 41, 433.

[8] T. Caillat, P. Ball, *MRS Bulletin Energy Quarterly*, **2013**, June 11.

[9] D.M. Rowe, Thermoelectric Handbook:macro to Nano. Boca Raton, CRC Press, **2006**.

[10] K. Biswas, J. He, Q. Zhang, G. Wang, C. Uher, V. P. Dravid, M. Kanatzidis, *Nature* **2012**, 489, 414.

[11] Z. Ren, S. Chen, *Materials Today.* **2013,** 16, 387.

[12] Y. Ma, R. Heijl, A.E.C. Palmqvist, *J Mater Sci.* **2013**, 48, 2767.

[13] T. Graf, C. Felser, S. Parkin, *Progress in Solid State Chemistry* **2011**, 39, 1.

[14] S. Ouardi, G.H. Fecher, B. Balke, X. Kozina, G. Stryganyuk, C. Felser, S. Lowitzer, D. Ködderitzsch, H. Ebert, E. Ikenaga, *Phys.Rev.B.* **2010**, 82, 085108.

[15] S. Populoh, M.H. Aguirre, O.C. Brunko, K. Galazhka, Y. Lu, A. Weidenkaff, *Scripta Mater.* **2012**, 66, 1073.

[16] D-Y. Jung, K. Kurosaki, C-E. Kim, H. Muta, S. Yamanaka, *J. Alloys Compd.* **2010**, 489, 328.

[17] C. Yu, T.-J. Zhu, R.-Z. Shi, Y. Zhang, X.B. Zhao, J. He, *Acta Materiala.* **2009**, **57**, 2757.

[18] K. Bartolomé, B. Balke, D. Zuckermann, M. Köhne, M. Müller, K. Tarantik, J. König, *J. Electron. Mater.* **2013**, 43, 1775.

[19] S. Ouardi, G.H. Fecher, B. Balke, M. Schwall, X. Kozina, G. Stryganyuk, C. Felser, E. Ikenaga, Y. Yamashita, S. Ueda, K. Kobayashi, *Appl. Phys. Lett.* **2010**, 97, 252113.

[20] L.D. Zhao, B.P. Zhang, J.F. Li, M. Zhou, W.S. Liu, *J. Alloy. Compd.* **2009**, 467, 91.

[21] E. Hatzikraniotis, K.T. Zorbas, I. Samaras, Th. Kyratsi, K.M. Paraskevopoulos, *J.Electron. Mater*. **2010**, 39, 2112.

[22] S.R. Culp, S.J. Poon, N. Hickman, T.M. Tritt, J. Blumm, *Appl. Phys. Lett.* **2006**, 88, 042106.

[23] H.J. Goldsmid, A.W. Penn, *Phys. Lett A.* **1968**, 27, 523.

[24] H. Xie, H. Wang Y. Pei, C. Fu, X. Liu, G.J. Snyder, X. Zhao, T. Zhu, *Adv. Funct. Mater.* **2013**, 23, 5123.